\documentclass[floatfix,aps,pre,amssymb]{revtex4}

\usepackage{graphicx}
\usepackage{mathptmx}

\begin{document}

\title{\textbf{Growth rate and the cutoff wavelength
of the Darrieus-Landau instability in laser ablation}}

\author{Mikhail Modestov, Vitaly Bychkov, Damir Valiev and Mattias Marklund}

\affiliation{Department of Physics, Ume{\aa} University, 901 87
Ume{\aa}, Sweden}

\bigskip

\begin{abstract}
The main characteristics of the linear Darrieus-Landau instability in
the laser ablation flow are investigated. The dispersion relation of
the instability is found numerically as a solution to an eigenvalue
stability problem, taking into account the continuous structure of the
flow. The results are compared to the classical Darrieus-Landau
instability of a usual slow flame. The difference between the two cases
is due to the specific features of laser ablation: high plasma
compression and strong temperature dependence of electron thermal
conduction. It is demonstrated that the Darrieus-Landau instability
in laser ablation is much stronger than in the classical case.
In particular, the maximum growth rate in the case of laser
ablation is about three times larger than that for slow flames. The
characteristic length scale of the Darrieus-Landau instability in
the ablation flow is comparable to the total distance from the
ablation zone to the critical zone of laser light absorption.
The possibility of experimental observations of the Darrieus-Landau
instability in laser ablation is discussed.
\end{abstract}

\maketitle

\section{Introduction}

Inertial confinement fusion (ICF) is believed to be one of the
promising energy sources in the 21$^{\rm st}$ century. The aim of
ICF is to compress a plasma target to densities and temperatures high
enough to trigger a thermonuclear reaction. Despite a great
technical development and progress in power supplies during the last decades \cite{Dunne-06},
hydrodynamic instabilities remain the limiting factor in fusion
performance and efficiency. In this respect, the most difficult obstacle in achieving
ICF is the Rayleigh-Taylor (RT) instability, which arises because of
target acceleration \cite{Eliezer,Bodner, Kull-91, Bychkov-94, Betti-95,
Betti-06,Sanz et al.,Bychkov_RTI_ICF-07}. Still, the RT instability is not the only
instability of importance in ICF; for example, the laser generated plasma is also subject
to the so called Darrieus-Landau (DL) instability 
\cite{Bychkov-94, Piriz-01,Piriz-03, Keskinen et al., Gotchev et
al., Goncharov et al., Modestov_RTI,Clavin-04,Bychkov-08}. Traditionally, the DL
instability is known as the hydrodynamic instability of a slow flame
\cite{LandauL, Zeldovich, review,Searby and Clavin}, which makes the flame front corrugated and increases the burning rate. Flame is the most typical example
of a deflagration wave, which is a front propagating due to energy
release and thermal conduction. The ablation flow generated by laser
radiation on a plasma target is also a deflagration wave
\cite{Manheimer, Fabbro,Bychkov_RTI_ICF-07}. Other interesting examples of deflagration
come from astrophysical applications like the big bang model and
 type I supernovae \cite{Link-92,Kamionkowski-Freese-92,Fragile-Anninos-03,Bychkov_Astron-95, Gamezo et al.2003,Bychkov_AstRep-06}. On the basis of similar physical properties of deflagrations, one should expect
the DL instability to develop in laser ablation. There has been much interest in the DL instability in ablation flows, see e.g. Refs.\
\cite{Bychkov-94, Piriz-01,Piriz-03, Keskinen et al., Gotchev et
al., Goncharov et al., Modestov_RTI,Clavin-04,Bychkov-08}. However, the most important features of the instability has remained
unclear up until now. The purpose of the present theoretical work is to
answer some of the key questions concerning the DL instability in laser
ablation and to indicate conditions for experimental observation of
the instability.

The classical theory of the DL instability considers an
incompressible flow generated by a slow flame \cite{LandauL,
Zeldovich, review, Searby and Clavin}. However, the parameters of
laser ablation flows in ICF are markedly different from those of slow
combustion. One of the most distinctive features of deflagration in
ICF is that the plasma velocity reaches the isothermal sound speed at
the critical surface of laser light absorption. For this reason,
laser ablation corresponds to the so-called Chapman-Jouguet (CJ)
deflagration, which is the fastest propagation regime
possible for a deflagration front. Thus, in order to describe the DL
instability in laser ablation one has to take into account strong
compression of the plasma flow with relatively high local values of
the Mach number. Another important property inherent to the plasma
flow is the strong temperature dependence of electron thermal
conduction. Electron thermal conduction determines the total thickness
and the internal structure of laser deflagration. It is expected
that the strong plasma compression and the electron thermal conduction properties will
significantly influence the properties of the DL instability in a laser generated plasma,
making it markedly different from the case of slow flames. A number
of papers was devoted to the linear and nonlinear stages of the DL
instability in ICF \cite{Piriz-03, Clavin-04, Sanz et al.,
Bychkov-08, Keskinen et al., Gotchev et al., Goncharov et al.}.
Still, these papers did not answer the most important questions
concerning the DL instability in laser ablation. In particular, we
would here like to outline the following three questions within the
problem:

\begin{enumerate}
\item How strong is the DL instability in laser ablation in comparison to the
classical case, i.e., is it stronger or weaker than the DL instability developing
at a slow flame front?
\item What is the characteristic length scale of the instability development?
This question is especially important from the experimental point of view,
since the answer determines the target size, for which the DL instability may be
observed.
\item What is the outcome of the DL instability at the nonlinear stage?
\end{enumerate}

There have been numerous attempts to answer the first question
within the model of a discontinuous deflagration front in a
compressible gas/plasma flow \cite{Kadowaki, He, Piriz-03, Piriz-01,
Bychkov-08}. Unfortunately, the discontinuous model encounters the
deficit of matching conditions at the deflagration front; the number
of unknown variables exceeds the number of conservation laws at the
front by one. The problem was encountered first within the studies
of the RT instability in laser ablation, see Refs.\ \cite{Bodner,
Bychkov-94, Bychkov_RTI_ICF-07}; more detailed discussion on this
subject may be found in Ref.\ \cite{Bychkov-08}. In the classical
incompressible limit of the DL instability at a slow flame front,
the extra condition is given by the so-called DL condition of a
constant deflagration speed with respect to the cold gas. This
condition may be proven rigorously, see Ref.\ \cite{Searby and Clavin,
review}. A counterpart of the DL condition for a compressible flow is not obvious and the
solution to the problem turned out to be sensitive to this assumed extra
condition, so that the different analytical theories proposed led
to qualitatively different results. For this reason, it was
unclear if strong plasma compression in laser ablation made the DL
instability stronger, or weaker, or only produces minor changes in the
instability strength. Within this context, Refs.\ \cite{Travnikov-97,
Travnikov-99} deserve special attention, since these papers
considered influence of gas compression on the DL instability taking
into account a continuous structure of the deflagration front.
Though Refs.\ \cite{Travnikov-97, Travnikov-99} were devoted to normal
combustion, and the respective results cannot be extrapolated
directly to the DL instability in laser ablation, the method
used in these papers may still be used for the ablation studies.
This method eliminates the deficit of boundary conditions and allows
for investigating the properties of the DL instability in laser ablation.
Here we employ the methods of Refs.\ \cite{Travnikov-97,
Travnikov-99} to study the linear stage of the DL instability in
a laser plasma, thus answering questions 1 and 2 of those outlined
above.

In the present paper we investigate main characteristics of the linear
DL instability in the laser ablation flow. We find the
dispersion relation of the instability numerically as a solution to an
eigenvalue stability problem taking into account the continuous structure of the
flow. We compare the results to the classical DL instability of a usual slow
flame. We show that difference between the two cases is due to two specific features
of laser ablation: a high plasma compression and a strong temperature dependence
of the electron thermal conduction. We further demonstrate that the DL
instability in laser ablation is much stronger than in the classical case.
In particular, the maximal growth rate of perturbations in laser ablation is
about three times larger than for slow flames. The characteristic length
scale of the DL instability in the ablation flow is comparable
to the total distance from the ablation zone to the critical zone of laser
light absorption. We discuss the possibility of experimental observations of
the DL instability in laser ablation.

\section{Basic equations and the stationary solution}

We describe the ablation plasma flow using hydrodynamic equations of
mass, momentum and energy conservation
\begin{equation}
\label{eq1}
{\frac{{\partial \rho}} {{\partial t}}} + \nabla \cdot (\rho {\rm {\bf u}})
= 0,
\end{equation}

\begin{equation}
\label{eq2}
\rho {\frac{{\partial {\rm {\bf u}}}}{{\partial t}}} + (\rho {\rm {\bf u}}
\cdot \nabla ){\rm {\bf u}} + \nabla P = 0,
\end{equation}

\begin{equation}
\label{eq3}
{\frac{{\partial}} {{\partial t}}}\left( {\rho C_{V} T +
{\frac{{1}}{{2}}}\rho u^{2}} \right) + \nabla \cdot {\left[ {\rho {\rm {\bf
u}}\left( {C_{P} T + {\frac{{1}}{{2}}}u^{2}} \right) - \kappa \nabla T}
\right]} = \Omega _{R} ,
\end{equation}

\noindent
and the equation of state of an ideal gas

\begin{equation}
\label{eq4}
P = {\frac{{\gamma - 1}}{{\gamma}} }C_{P} \rho T,
\end{equation}

\noindent where $\gamma = 5/3$ is the adiabatic exponent, $C_P$ and
$C_V$ are the heat capacities at constant pressure and volume
respectively. Electron thermal conduction $\kappa $ depends on
temperature as $\kappa = \kappa_c (T/T_c)^{5/2}$, where label ``c''
refers to the critical surface of laser light absorption. Laser
light absorption brings energy into the plasma and, together with
thermal conduction, it drives the flow. Absorption takes place when
plasma frequency is equal to the laser frequency

\begin{equation}
\label{eq5}
\omega ^{2} = \omega _{p}^{2} = {\frac{{4\pi e^{2}\rho _{c}}} {{m_{e}
M}}},
\end{equation}

\noindent which determines plasma density $\rho _{c} $ at the
critical surface (here $M$ is plasma mass per one electron).
Decrease of the laser light intensity due to absorption may be
described as \cite{Eliezer}

\begin{equation}
\label{eq6} \frac{dI}{dz} = KI,
\end{equation}

\noindent with the absorption coefficient

\begin{equation}
\label{eq7} K \propto \frac{\rho ^2}{T^{3/2}} \left( {1 -
{\frac{\rho} {\rho _c} }} \right)^{-1/2}.
\end{equation}

\noindent The absorption coefficient diverges at the critical
surface, and, therefore, the process of energy absorption is
strongly localized at the surface. In the studies of ablation flow
and the RT and DL instabilities in the flow, the energy release is
typically presented by $\delta $-function \cite{Bodner, Manheimer,
Bychkov-94,Piriz-01,Piriz-03}. Such replacement is possible since the instabilities
develop on the length scales much larger than the region of energy
release and involve bending of this region as a whole without
changing its internal structure. In the present paper we solve the
problem of the DL instability numerically. In the numerical
solution, it is more convenient to imitate the $\delta $-function by
a transitional zone of finite width determined by some continuous
function $\Omega _R$ of energy gain in the flow included into Eq.
(\ref{eq3}). Here, we chose the function in the form suggested in
Ref. \cite{Bychkov-08}

\begin{equation}
\label{eq8} \Omega _{R} = \Omega \left( {\rho - \rho _{c}}
\right)^{n}\exp ( - \beta \rho/\rho_c ).
\end{equation}

\noindent The function $\Omega _{R} $ was constructed taken into
account similarity with the Arrhenius law in combustion, where
$\beta $ plays the role of the scaled activation energy and $n$ is
similar to the reaction order. It is well-known that the Arrhenius
law provides strong localization of energy release for any
reasonable $n$ in the case of sufficiently large $\beta > > 1$, see
\cite{Zeldovich, review}. In combustion science the Arrhenius
reaction is sensitive to temperature changes \cite{Zeldovich,
review}. Here we construct the function of energy gain $\Omega _{R}
$ sensitive to density variations, as it takes place in laser
ablation. Choosing large parameter $\beta \to \infty $ we obtain
energy gain strongly localized at the critical surface with $\rho
\to \rho _{c} $. On the other hand, the numerical solution demands a
finite width of the zone of energy release and the finite value of
the parameter $\beta $. In most of our calculations we use $\beta =
90$, $n = 2$. We also investigate sensitivity of the physical
results to the choice of these parameters. We demonstrate that these
parameters have minor influence upon the properties of the DL
instability and at high values of $\beta$ this influence vanishes.
The function $\Omega _{R} $ given by Eq. (\ref{eq8}) allows a planar
stationary solution consisting of two uniform flows of cold heavy
plasma (label ``a'') and hot light plasma (label ``c'') separated by
a transitional region, which is the deflagration front. The labels
``a'' and ``c'' originate from the ablation and critical surfaces in
the laser deflagration. Typical internal structure of the
deflagration front is illustrated in Fig. 1. The described geometry
is common in the theoretical studies of the RT and DL instabilities
of the ablation flow \cite{Bodner, Bychkov-94, Piriz-01, Piriz-03,
He,Betti-95,Kull-91,Sanz et al.,Bychkov_RTI_ICF-07}, though it does
not take into account the rarefaction wave in the hot light plasma
beyond the critical surface. As the main advantage of such a choice,
we may consider different values of the Mach number in the light
plasma, which would be impossible with a rarefaction wave. Changing
the Mach number, we can go over continuously from the case of
classical incompressible DL instability of a usual flame front to
the case of laser ablation with strong influence of plasma
compression. We stress that the main purpose of the present paper is
to compare the DL instability in laser ablation to the classical
case. Therefore, in general, we will discuss stability of a
deflagration front, which covers the cases of usual flames (slow
combustion) and laser ablation as two asymptotic limits of
negligible compression effects and ultimately strong plasma
compression.

We start our analysis with describing internal structure of the
stationary planar deflagration (laser ablation) flow; this is the
first step in the stability analysis. Figure 1 illustrates plasma
density, temperature and energy release in the flow obtained
numerically as described below. The deflagration front in Fig. 1
propagates to the left with constant velocity $U_{a} $ (the ablation
velocity) in the negative direction of z-axis. Velocity of a usual
flame is determined by the rate of energy release and thermal
conduction. Ablation velocity is determined by the critical density
and the laser light intensity \cite{Manheimer}. We adopt the
reference frame of the deflagration front. In that case the front is
at rest, but the cold heavy plasma flows to the right with uniform
velocity $u_{z} = U_{a} $, undergoes transition in density and
temperature in the deflagration wave, and, finally, the hot light
plasma gets drifted away with uniform velocity $u_{z} = U_{c} $.

Equations of mass and momentum transfer (\ref{eq1}), (\ref{eq2}) may be integrated for the
planar stationary deflagration as

\begin{equation}
\label{eq9}
\rho u_{z} = \rho _{a} U_{a} = \rho _{c} U_{c} ,
\end{equation}

\begin{equation}
\label{eq10}
P + \rho u_{z}^{2} = P_{a} + \rho _{a} U_{a}^{2} = P_{c} + \rho _{c}
U_{c}^{2} .
\end{equation}

\noindent One of the main dimensionless parameters in the problem is
the expansion factor

\begin{equation}
\label{eq11} \Theta = \frac{\rho _a}{\rho_c} = \frac{U_c}{U_a},
\end{equation}

\noindent which shows density drop from the original cold plasma to
the critical surface. Both in flames and laser ablation, the
expansion factor is rather large, $\Theta = 5 - 10$, see
\cite{review, Betti-06}. In the case of ablation flow, the laser
light frequency determines the critical density and the expansion
factor. The other important parameter is the Mach number in the
light plasma (gas) corresponding to the adiabatic sound

\begin{equation}
\label{eq12} M\!a_{c} = U_{c} \sqrt {\frac{\rho_c}{\gamma P_c} } .
\end{equation}

\noindent The Mach number is negligible in the classical case of
usual flames, and it may be taken zero with a very good accuracy. On
the contrary, laser ablation provides an ultimately large value of
the Mach number possible for a deflagration flow. In laser ablation,
the isothermal Mach number is equal unity $\rho _{c} U_{c}^{2} /
P_{c} = 1$ in the light plasma, and we have $M\!a_{c}^{2} = 1 /
\gamma $ for the adiabatic Mach number. In the present work we
consider a general case of a deflagration front with an arbitrary
Mach number changing within the limits $0 \le M\!a_{c}^{2} < 1 /
\gamma $. The internal structure of the deflagration front follows
from the stationary equation of energy transfer

\begin{equation}
\label{eq13}
\frac{d}{dz}{\left[ {\rho _{c} U_{c} \left( {C_{P} T +
\frac{1}{2}u^{2}} \right) - \kappa \frac{dT}{dz}} \right]} = \Omega
_{R} .
\end{equation}

\noindent Characteristic width of the front is determined by thermal
conduction in the hot region

\begin{equation}
\label{eq14} L_{c} \equiv {\frac{\kappa_c} {C_p \rho_c U_c}}.
\end{equation}

\noindent The problem involves one more parameter of length
dimension

\begin{equation}
\label{eq15} L_{a} \equiv {\frac{\kappa_a} {C_p \rho_a U_a} } =
{\frac{\kappa_a} {C_p \rho_c U_c} }
\end{equation}

\noindent related to thermal conduction in the cold flow. Because of
the strong temperature dependence of electron thermal conduction,
these two length scales are quite different $L_{c} / L_{a} = (T_{c}
/ T_{a} )^{5 / 2}$. For example, for the temperatures ratio $T_{c} /
T_{a} = 6$, the length scales differ by two orders of magnitude
$L_{c} / L_{a} \approx 88$. The strong difference in these length
scales is one of the specific features of laser ablation in
comparison with usual flames.

We introduce dimensionless variables for plasma density, temperature,
velocity and coordinate

\begin{equation}
\label{eq16}
\varphi = {\frac{{\rho}} {{\rho _{c}}} },
\quad
\theta = {\frac{{T}}{{T_{c}}} },
\quad
u = {\frac{{u_{z}}} {{U_{c}}} },
\quad
\xi = {\frac{{z}}{{L_{c}}} }.
\end{equation}

\noindent Then we can rewrite Eq. (\ref{eq13}) as

\begin{equation}
\label{eq17} {\frac{\partial} {\partial \xi} }{\left[ {\theta +
{\frac{{(\gamma - 1)}}{{2\varphi ^{2}}}}M\!a_{c}^{2} - \theta ^{5 /
2}{\frac{{\partial \theta }}{{\partial \xi}} }} \right]} = \Lambda
\left( {\varphi - 1} \right)^{n}\exp \left( { - \beta \varphi}
\right) ,
\end{equation}

\noindent where $\Lambda = L_{c}^{2} \Omega \rho _{c}^{n} \left(
{\kappa _{c} T_{c}} \right)^{ - 1}$ is an eigenvalue of the
stationary problem. The relation between temperature and density in
a deflagration flow follows from Eqs. (\ref{eq9}), (\ref{eq10})

\begin{equation}
\label{eq18} \theta = {\frac{{1 + \gamma M\!a_{c}^{2}}} {{\varphi}}
} - {\frac{{\gamma M\!a_{c}^{2}}} {{\varphi ^{2}}}} .
\end{equation}

\noindent In the incompressible limit of usual flames, $M\!a_{c}^{2}
< < 1$, this relation is reduced simply to $\varphi \theta = 1$, so
that temperature ratio $T_{c} / T_{a} $ is determined by the
expansion ratio, $T_{c} / T_{a} = \Theta $. In the case of strong
gas compression, these two values differ considerably. For example,
in the case of laser ablation with the critical Mach number
$M\!a_{c}^{2} = 1 / \gamma $, we find from Eq. (\ref{eq18}) that

\begin{equation}
\label{eq19} \frac{T_c} {T_a} = {\frac{\Theta^2} {2 \Theta - 1}} .
\end{equation}

\noindent For high values of density drop, $2\Theta > > 1$,
temperature ratio is about twice smaller than the expansion factor,
$T_{c} / T_{a} \approx \Theta / 2$. Because of the reduced
temperature ratio, the effect of two different length scales Eqs.
(\ref{eq14}), (\ref{eq15}) in the ablation flow is expected to
become weaker than in a similar incompressible flow. With
temperature ratio changing from $T_{c} / T_{a} = \Theta $ in the
incompressible case to $T_{c} / T_{a} \approx \Theta / 2$ in the
ablation flow we find the ratio of length scales $L_{c} / L_{a} =
(T_{c} / T_{a} )^{5 / 2}$ decreasing by the factor of
$(\ref{eq2})^{5 / 2} \approx 5.7$. For this reason, the profiles of
density and temperature are expected to be much smoother in the
ablation flow in comparison with an incompressible counterpart.

We solve Eq.(\ref{eq17}) together with Eq. (\ref{eq18}) numerically
for different values of the Mach number. Typical solution to the
problem for incompressible case is shown in Fig. 1. First of all, we
have to make sure, that the ablation front structure is not
sensitive to our choice of the energy gain function $\Omega _{R} $.
We investigate dependence of the density and temperature profiles on
the parameter $\beta $. Figure 2 shows density and energy release
for $\Theta = 6$, $n = 2$, $M\!a_{c} = 0.5$, $\beta =
20;\;60;\;140$. Density profiles for these three values of $\beta $
coincide almost everywhere except for the zone of energy gain.
Still, even inside this zone, the difference between the density
profiles is minimal, and cannot be recognized on Fig. 2. Similar
result holds for the temperature profiles. Thus, we can conclude
that parameter $\beta $ does not affect the density and temperature
profiles in the ablation front. At the same time, parameter $\beta $
influences strongly the profile of energy release itself: the
respective plots become much more localized with $\beta $
increasing. Thickness of the critical surface should be zero in the
hydrodynamic description of the ablation flow. Therefore, finite
width of the zone of energy gain implies certain inaccuracy of the
numerical solution. The level of inaccuracy of the model may be
evaluated as the characteristic width of the energy gain zone in the
dimensionless variables (scaled by the total thickness of the
deflagration front $L_{c} )$. As we can see in Fig. 2, half width of
the energy peak is rather wide for $\beta = 20$; it takes more than
20\% of the whole deflagration front. As we increase $\beta $, this
width becomes smaller and for $\beta \ge 90$ the inaccuracy is less
than 5\% . The other parameter of the energy gain function in Eq.
(\ref{eq8}), $n$, affects the shape of the energy release peak only
slightly even for moderate values of $\beta $. For $\beta = 90$ the
parameter $n$ has a negligible effect on the physical results. Thus,
Eq. (\ref{eq8}) may imitate the energy gain in the ablation flow
quite well; in all following numerical solutions and figures $\beta
= 90$ and $n = 2$.

As the next step, we solve Eq. (\ref{eq17}) for different values of
the Mach number.  Figure 3 shows profiles of density and energy
release, for $M\!a_{c} = 0;\,0.5;\,0.75$, other parameters being
fixed as $\Theta = 6$, $\beta = 90$, $n = 2$; Figure 4 presents the
respective temperature profiles. In the case of incompressible flow,
$M\!a_{c} = 0$, the density profile demonstrates clearly the effect
of two length scales, Eqs. (\ref{eq14}), (\ref{eq15}), produced by
the temperature-dependent thermal conduction. The profile is rather
smooth in the hot region close to the critical surface, and it
becomes sharp in the cold plasma close to the ablation surface with
large density gradient. In fact, it is this property, which allows
distinguishing two effective surfaces in the flow: the ablation and
critical surfaces \cite{Manheimer, Piriz-01, Bychkov-94}. In the
case of strong compression with $M\!a_{c} = 0.75$, the density
profile becomes much smoother. Smoothing of the density profile
concerns the region of cold plasma mainly. Another specific feature
of the density profile at high values of the Mach number is shown at
the insert of Fig. 3. With the Mach number approaching the maximal
possible value $M\!a_{c}^{2} = 1 / \gamma $, we observe development
of another mini-region of relatively high density gradient close to
the critical surface. This effect may be also obtained analytically
from Eq. (\ref{eq18}). Close to the critical surface, expanding
density and temperature in power series with respect to $\varphi - 1
< < 1$, $1 - \theta < < 1$, we obtain the relation $\varphi - 1
\approx \sqrt {1 - \theta} $, or $\rho / \rho _{c} - 1 \approx \sqrt
{1 - T / T_{c}}  $ in the dimensional values. Taking energy gain in
the form of $\delta (z)$-function, we have temperature achieving the
final value $T_{c} $ at finite point $z = 0$ smoothly. This leads to
the square-root singularity in the density gradient. We can observe
the trace of such a singularity in the insert of Fig. 3, though
smoothed because of the finite width of the energy gain zone. We
also observe changing shape of the energy gain with Mach number. The
energy gain zone becomes much thinner at high values of the Mach
number. This happens because the energy release Eq. (\ref{eq8}) is
sensitive to the density profiles. Sharp gradients of density at
$M\!a_{c}^{2} = 1 / \gamma $ make the zone of energy gain sharper as
well. As a result, the model Eq. (\ref{eq8}) works much better at
high values of the Mach number corresponding to the laser ablation
flow. Figure 4 demonstrates that temperature profiles become also
smoother with increasing Mach number, which is similar to density
profiles. Besides, temperature ratio at the ablation and critical
surfaces decreases with increasing the Mach number, as we
demonstrated in Eqs. (\ref{eq18}), (\ref{eq19}). The numerical
solution for the planar stationary flow illustrated in Figs. 1-4
provides the basis for the stability analysis performed in the next
section.

\section{Linearized equations}

We solve the stability problem for small perturbations of any value $\phi $
in the form:

\begin{equation}
\label{eq20}
\phi (x,z,t) = \phi  (z) + \tilde {\phi} (z)\exp (\sigma t + ikx){\rm
,}
\end{equation}

\noindent where the first term in the right-hand side of
(\ref{eq20}) stands for the stationary flow, the second term
describes linear perturbations, $\sigma $ is the instability growth
rate and $k = 2\pi / \lambda $ is the perturbation wave number. In
general, $\sigma $ may have both a real part (growth rate) and an
imaginary part (frequency). However, in the case of the DL
instability $\sigma $ is only real; the instability develops when
$\sigma $ is positive.

The linearized system (1-3) takes the form

\begin{equation}
\label{eq21}
\sigma \tilde {\rho}  + {\frac{{d}}{{dz}}}\left( {\rho \tilde {u}_{z} +
\tilde {\rho} u_{z}}  \right) + ik\rho \tilde {u}_{x} = 0,
\end{equation}

\begin{equation}
\label{eq22} \sigma \rho \tilde {u}_{x} + \rho u_{z} {\frac{{d\tilde
{u}_{x}}} {{dz}}} + ik\widetilde {P} = 0,
\end{equation}


\begin{equation}
\label{eq23} \sigma \rho \tilde {u}_{z} + \rho u_{z} {\frac{{d\tilde
{u}_{z}}} {{dz}}} + {\frac{{du_{z}}} {{dz}}}\left( {\tilde {\rho}
u_{z} + \rho \tilde {u}_{z}} \right) + {\frac{{d\widetilde
{P}}}{{dz}}} = 0,
\end{equation}

\[
\sigma (\rho C_{V} \widetilde {T} + \rho u\tilde {u} - \tilde {\rho}
RT) + \left( {\tilde {\rho} u_{z} + \rho \tilde {u}_{z}}
\right){\frac{{\partial }}{{\partial z}}}\left( {C_{P} T +
{\frac{{1}}{{2}}}u^{2}} \right) + \rho u_{z} {\frac{{\partial}}
{{\partial z}}}\left( {C_{P} \widetilde {T} + u\tilde {u}} \right)
\]

\begin{equation}
\label{eq24} \quad + \frac {\kappa_c} {{T_c}^{5 / 2}} \left[ k^2
T^{5 / 2} \widetilde T - \frac{\partial} {\partial z} \left( \frac
{5} {2} \frac{\partial T}{\partial z} \widetilde T + T \frac{
\partial \widetilde T} {\partial z} \right) \right] - \widetilde
\Omega_R = 0.
\end{equation}

\noindent Similar to \cite{Bychkov-94}, we introduce the
dimensionless perturbations of mass flow $\tilde {j}$, transverse
velocity $\tilde {v}$, temperature $\tilde {\theta} $ and dynamic
pressure $\widetilde {\Pi} $ as

\begin{equation}
\label{eq25}
\tilde j = {\frac{{\rho \tilde {u}_{z} + \tilde {\rho}
u_{z}}} {{\rho _{c} u_{c}}} },
\quad \tilde {v} = {\frac{{i\tilde
{u}_{x}}} {{u_{c}}} },
\quad \tilde {\theta}  = {\frac{{\widetilde
{T}}}{{T_{c}}} },
\quad \widetilde {\Pi}  = {\frac{{\widetilde {P} +
\tilde {\rho} u_{z}^{2} + 2\rho u_{z} \tilde {u}_{z}}} {{\rho _{c}
u_{c}^{2}}} }
\end{equation}

\noindent with the scaled wave number and the perturbation growth
rate

\begin{equation}
\label{eq26} {\rm K} = kL_c , \quad S = \frac{\sigma L_c}{U_c}.
\end{equation}

\noindent Then the linearized system (\ref{eq21})-(\ref{eq24}) is

\begin{equation}
\label{eq27} {\frac{{d\tilde {j}}}{{d\xi}} } = 2S{\frac{{\gamma
M\!a_{c}^{2} u}}{{w}}}\tilde {j} - {\rm K}\varphi \tilde {v} -
S{\frac{ \gamma M\!a_{c}^{2} } {w}}\widetilde {\Pi}  +
S{\frac{\varphi} {w}}\tilde {\theta} ,
\end{equation}

\begin{equation}
\label{eq28} {\frac{{d\tilde {v}}}{{d\xi}} } = - 2{\rm
K}{\frac{{\theta u}}{{w}}}\tilde {j} - S\varphi \tilde {v} + {\rm
K}{\frac{{\theta}} {{w}}}\widetilde {\Pi}  - {\rm
K}{\frac{{u}}{{w}}}\tilde {\theta} ,
\end{equation}

\begin{equation}
\label{eq29} {\frac{{d\widetilde {\Pi}} }{{d\xi}} } = - S\tilde {j}
- {\rm K}\tilde {v}{\rm} ,
\end{equation}

\[
\theta ^{5 / 2}{\frac{{d^{2}\tilde {\theta}} }{{d\xi ^{2}}}} + \theta ^{3 /
2}\left( {5{\frac{{d\theta}} {{d\xi}} } - A_{\psi}  \theta}
\right){\frac{{d\tilde {\theta}} }{{d\xi}} } +
\]

\begin{equation}
\label{eq30}
 + {\frac{{5}}{{2}}}\theta ^{1 / 2}{\left[ {{\frac{{3}}{{2}}}\left(
{{\frac{{d\theta}} {{d\xi}} }} \right)^{2} + \theta
{\frac{{d^{2}\theta }}{{d\xi ^{2}}}} - A_{\psi}  \theta
{\frac{{d\theta}} {{d\xi}} }} \right]}\tilde {\theta}  = A_{j}
\tilde {j} + A_{v} \tilde {v} + A_{\Pi} \widetilde {\Pi}  +
A_{\theta} \tilde {\theta},
\end{equation}

\noindent where $w = \theta - \gamma M\!a_{c}^{2} u^{2}$ and the
coefficients in Eq. (\ref{eq30}) are

\begin{equation}
\label{eq31} A_{j} = {\frac{{\partial \theta}} {{\partial \xi}} } +
\left( {\gamma - 1} \right)M\!a^{2}{\left[ {{\frac{{u}}{{w}}}\left(
{S\left( {3\theta + \gamma M\!a^{2}u^{2}} \right){\frac{{\theta}}
{{w}}} + {\frac{{2\gamma}} {{\gamma - 1}}}{\frac{\partial \Omega
_{R}} {\partial \varphi}} } \right) + {\frac{\partial} {\partial
\xi} }{\left\{ {{\frac{{3\theta + \gamma
M\!a^{2}u^{2}}}{{2w}}}u^{2}} \right\}}} \right]},
\end{equation}

\begin{equation}
\label{eq32} A_{v} = (\gamma - 1)M\!a_{c}^{2} {\frac{{u}}{{w}}}{\rm
K}\theta ,
\end{equation}

\begin{equation}
\label{eq33} A_{\Pi}  = - (\gamma - 1)M\!a_{c}^{2} {\left[
{S{\frac{{\theta + \gamma M\!a_{c}^{2} u^{2}}}{{w^{2}}}}\theta +
{\frac{{\gamma}} {{(\gamma - 1)w}}}{\frac{{\partial \Omega _{R}}}
{{\partial \varphi}} } + \gamma M\!a_{c}^{2} {\frac{{\partial}}
{{\partial \xi}} }{\left\{ {{\frac{{u^{3}}}{{w}}}} \right\}}}
\right]}
\end{equation}

\[
A_{\theta}  = S{\frac{{\varphi \theta}} {{w}}} + {\rm K}^{2}\theta
^{5 / 2} + {\frac{\partial \Omega _{R}} {\partial
\varphi}}{\frac{{\varphi}} {{w}}} -
{\frac{{5}}{{2w}}}{\frac{{\partial \theta }}{{\partial \xi}} } +
\]

\begin{equation}
\label{eq34}
 + M\!a_{c}^{2} {\left[ {{\frac{{5u^{2}}}{{2w\theta}} }{\frac{{\partial \theta
}}{{\partial \xi}} } + S\left( {(2\gamma - 3)\theta + \gamma
M\!a_{c}^{2} u^{2}} \right){\frac{{u}}{{w^{2}}}} + (\gamma -
1){\frac{{\partial }}{{\partial \xi}} }{\left\{
{{\frac{{u^{2}}}{{w}}}} \right\}}} \right]},
\end{equation}

\begin{equation}
\label{eq35} A_{\psi}  = \theta ^{ - 5 / 2}{\frac{{\theta -
M\!a_{c}^{2} u^{2}}}{{w}}}{\rm }.
\end{equation}

\noindent As the boundary conditions to the system
(\ref{eq27})-(\ref{eq30}), we demand that perturbations vanish at
infinity in the uniform flows of cold plasma ahead of the ablation
zone and the hot plasma behind the critical zone of energy gain. The
coefficients in Eqs. (\ref{eq27})-(\ref{eq30}) are constant in the
uniform flows. This allows us writing down the perturbations in an
exponential form $\tilde {\phi} \left( {\xi}  \right) = \tilde
{\phi} \exp \left( {\mu \xi} \right)$, where $\mu > 0$ in the heavy
plasma ($\xi \to - \infty )$ and $\mu < 0$ in the light plasma
$\left( {\xi \to \infty}  \right)$. In some particular simplified
limits the structure of the perturbation modes in the uniform flows
may be written analytically, e.g. see \cite{Bychkov-08}. However, in
the present case we can do it only numerically, which becomes
another step in the numerical solution to the problem.

\section{The algorithm for the numerical solution}

The derived system (\ref{eq27})-(\ref{eq30}) is rather complicated and it may be solved only
numerically. We introduce an auxiliary variable $\psi = \theta ^{5 /
2}\partial \theta / \partial \xi $ to obtain two differential equations of
the first order instead of Eq. (\ref{eq30}) as

\begin{equation}
\label{eq36} {\frac{{d\tilde {\theta}} }{{d\xi}} } = -
{\frac{{5}}{{2}}}{\frac{{1}}{{\theta}} }{\frac{{d\theta}} {{d\xi}}
}\tilde {\theta}  +  \theta ^{ - 5 / 2} \widetilde {\psi},
\end{equation}

\begin{equation}
\label{eq37} {\frac{{\partial \widetilde {\psi}} }{{\partial \xi}} }
= A_{j} \tilde {j} + A_{v} \tilde {v} + A_{\Pi}  \widetilde {\Pi}  +
A_{\theta}  \tilde {\theta}  + A_{\psi}  \widetilde {\psi} .
\end{equation}

\noindent Thus we have a system of five differential equations of
the first order with the scaled growth rate $S$ as an eigenvalue.
The purpose of the solution is to find the dispersion relation $S =
S\left( {{\rm K}} \right)$. The system may be written in a matrix
form

\begin{equation}
\label{eq38}
{\frac{{\partial \tilde {\phi}} }{{\partial \xi}} } = {\rm {\bf F}}\tilde
{\phi} ,
\end{equation}

\noindent where $\tilde {\phi} $ is a vector of the perturbations
and ${\rm {\bf F}}$ is the matrix

\begin{equation}
\label{eq39}  {\rm {\bf F}} = {\left[ {{\begin{array}{*{20}c}
 {2S {\frac{{\gamma M\!a_{c}^{2} u}}{{w}}}} \hfill & { - {\rm K}\varphi}
\hfill & { - S {\frac{{\gamma M\!a_{c}^{2}}} {{w}}}} \hfill & {S
{\frac{{\varphi}} {{w}}}} \hfill & {0} \hfill \\
 { - 2{\rm K}{\frac{{\theta u}}{{w}}}} \hfill & { - S \varphi}  \hfill
& {{\rm K}{\frac{{u}}{{w}}}} \hfill & {{\rm K}{\frac{{\theta}} {{w}}}}
\hfill & {0} \hfill \\
 { - S}  \hfill & { - {\rm K}} \hfill & {0} \hfill & {0} \hfill & {0}
\hfill \\
 {0} \hfill & {0} \hfill & {0} \hfill & { -
{\frac{{5}}{{2}}}{\frac{{1}}{{\theta}} }{\frac{{d\theta}} {{d\xi}} }} \hfill
& {\theta ^{ - 5 / 2}} \hfill \\
 {A_{j}}  \hfill & {A_{v}}  \hfill & {A_{\Pi}}   \hfill & {A_{\theta}}
\hfill & {A_{\psi}}   \hfill \\
\end{array}}}  \right]}.
\end{equation}

\noindent The numerical solution consists of the following steps.
First we look for the density profile in the stationary solution to
Eq. (\ref{eq17}). Equation (\ref{eq17}) is integrated numerically
from the uniform flows of cold and hot plasma to the central parts
of the deflagration transitional region. Both solutions are matched
at a certain density value; we checked that the physical results do
not depend on the choice of the matching point. When density
distribution is known, temperature and velocity profiles are
determined using Eqs. (\ref{eq9}) and (\ref{eq18}). The respective
numerical problem involves several length scales, which creates an
additional difficulty in the solution. In the case of highly
compressible flow the peak of energy release is extremely sharp; but
we have to resolve it properly, since it is used afterwards in the
solution to the stability problem. As the next step, we find modes
$\tilde {\phi} \left( {\xi}  \right) = \tilde {\phi} \exp \left(
{\mu \xi}  \right)$ in the uniform flows, which determine the
boundary conditions for Eqs. (\ref{eq27}) -- (\ref{eq29}). In the
uniform flows, Eq. (\ref{eq38}) is a system of linear ordinary
equations with constant coefficients. In order to find the factors
$\mu $, we solve

\begin{equation}
\label{eq40}
{\left| {{\rm {\bf F}} - {\rm {\bf {E}}}\mu}  \right|} = 0,
\end{equation}

\noindent where ${\rm {\bf {E}}}$ is the unit matrix. Thus we
obtain an equation for $\mu $ in the form of a polynomial of the
fifth order. In the incompressible case this equation may be solved
analytically. Five different roots correspond to the vorticity mode,
two sound modes and two modes of thermal conduction and/or energy
gain \cite{Liberman-94,Bychkov-94,Bychkov-08}. Solving Eq.
(\ref{eq40}) numerically we find five modes with two positive and
three negative values taking $\mu > 0$ for $\xi \to - \infty $ and
$\mu < 0$ for $\xi \to \infty $. Finally, we integrate the system
(\ref{eq38}) numerically in the transitional region of the
deflagration flow. We perform the numerical integration two times
from the right-hind side and three times from the left-hand side
with boundary conditions determined by different modes. We match
these five solutions at a certain point between the maximum of the
energy release and the ablation zone of sharp density gradients.
Again, the physical results do not depend on the choice of the
matching point. Then we obtain a matrix consisting of twenty five
values describing flow perturbations for five modes. Taking the
determinant of this matrix equal zero, we find the dispersion
relation$\,S = S\left( {{\rm K}} \right)$.

\section{Results and discussion}

Before presenting the numerical solution to the stability problem,
we explain the physical results we are looking for. In the classical
case of an infinitely thin flame front propagating in an
incompressible flow, the instability growth rate was obtained by
Darrieus and Landau \cite{LandauL, Zeldovich} as

\begin{equation}
\label{eq41}
\sigma = \Gamma U_{a} k,
\end{equation}

\noindent where the coefficient $\Gamma $ depends on the expansion
factor $\Theta $ only

\begin{equation}
\label{eq42}
\Gamma = {\frac{{\Theta}} {{\Theta + 1}}}\left( {\sqrt {\Theta + 1 - 1 /
\Theta}  - 1} \right).
\end{equation}

\noindent Traditionally the DL instability growth rate is scaled by
the front velocity $U_{a} $ instead of $U_{c} = \Theta U_{a} $ used
as the velocity unit in Secs. II, III. Here we follow the tradition
and use $U_{a} $ for scaling when presenting the results. The
approach of an infinitely thin front of the classical solution holds
for long wavelength perturbations $kL_{c} < < 1$. Taking into
account gas compression, we should expect the dispersion relation
for an infinitely thin deflagration front in the same form as Eq.
(\ref{eq41}), but with the coefficient $\Gamma $ depending on the
Mach number, $\Gamma = \Gamma (\Theta ,M\!a_{c} )$. More general
solution to the stability problem takes into account finite
thickness of the deflagration front. Finite thickness of the
deflagration front leads to stabilization of the DL instability at
sufficiently short wavelengths. In the case of usual slow flames of
finite thickness, the analytical solution to the problem may be
found e.g. in \cite{Searby and Clavin}. Written in the form of
Taylor expansion in relatively small perturbation wave number,
$kL_{c} < < 1$, the solution may be presented as

\begin{equation}
\label{eq43}
\sigma = \Gamma U_{a} k(1 - k / k_{cut} ),
\end{equation}

\noindent where $k_{cut} $ is the cut-off wave number, $\lambda
_{cut} = 2\pi / k_{cut} $ is the cut-off wavelength. The
approximation of the small cut-off wave number $k_{cut} L_{c} < < 1$
holds with reasonable accuracy for usual flames, e.g. see the review
\cite{review}. The cut-off wavelength is proportional to the
thickness of the deflagration front, $\lambda _{cut} \propto L_{c}
$. In the incompressible flow, the coefficient of proportionality
depends on the expansion factor $\Theta $ and on the type of thermal
conduction. Particularly, in the case of $\kappa \propto T^{5 / 2}$
and an incompressible flow, the theory \cite{Searby and Clavin}
predicts

\begin{equation}
\label{eq44} \frac {\lambda_{cut}} {L_c} =  {\frac{{4 \pi \Theta}}
{{\Theta - 1}}}
\left[{(1 - \Theta ^{ - 5 /
2}){\frac{{\Theta + 1}}{{5(\Theta - 1})}} + {\frac{{1}}{{7}}}} (1 - \Theta
^{ - 7 / 2})\right].
\end{equation}

\noindent For typical expansion factors $\Theta = 6 - 8$, Eq.
(\ref{eq44}) predicts the cut-off wavelength $\lambda _{cut} \approx
6L_{c} $. For comparison, the DL cut-off for usual flames is
considerably larger, being about $\lambda _{cut} \approx 20L_{c} $,
see \cite{review}. According to the dispersion relation
(\ref{eq43}), there is a maximum of the instability growth rate
$\sigma _{\max}  $ achieved at a certain finite perturbation
wavelength $\lambda _{\max}  $. Equation (\ref{eq43}) predicts the
wavelength of the maximum to be twice larger than the cut-off
wavelength, $\lambda _{\max}  = 2\lambda _{cut} $. Taking into
account gas compression, one should expect that all these values
depend on the Mach number, $M\!a_{c} $. Therefore, the purpose of
the present work is to investigate dependence of the parameters
$\Gamma $, $\lambda _{cut} $, $\lambda _{\max}  $, $\sigma _{\max} $
on the Mach number $M\!a_{c} $ changing within the interval $0 <
M\!a_{c} < 1/ \sqrt {\gamma} $. We demonstrate below that this
dependence is quite strong.

Figure 5 presents our numerical solution to Eqs.
(\ref{eq27})-(\ref{eq30}), that is the scaled instability growth
rate versus the perturbation wave number. The dispersion relation is
shown for different values of the Mach number, $M\!a_{c} =
0;\;0.5;\;0.65;\;0.7;\;0.73$, with the expansion factor $\Theta = 6$
and a strongly localized energy release for $\beta = 90$, $n = 2$,
see Sec. II. As we can see, the DL instability becomes much stronger
with increasing the Mach number; plasma (gas) compression provides a
strong destabilizing effect. This result looks very similar to one
obtained for a flame in a compressible flame in Refs.
\cite{Travnikov-97, Travnikov-99}. The destabilization concerns all
parameters of the instability: the factor $\Gamma $  in
Eq.(\ref{eq41}), the maximal instability growth rate, $\sigma
_{\max} $, and the cut-off wave number, $\lambda _{cut} $. These
results are presented in Figs. 6 -- 8, respectively. Figure 6 shows
the factor $\Gamma $, which determines strength of the DL
instability in the case of an infinitely thin front. The solution is
obtained for the expansion factors $\Theta = 6,\;10$. According to
Fig. 6, the DL instability for an infinitely thin ablation front
(with ultimately strong plasma compression) is almost twice stronger
than in the incompressible flow. The ratio of $\Gamma $-factors
obtained for $M\!a_{c} = 0.73$ and $M\!a_{c} = 0$ is about 1.8 for
$\Theta = 6, 10$. Some analytical theories for the DL instability in
compressible flows predicted also increase of the growth rate with
the Mach number \cite{Kadowaki, He}. Still, the numerical results
show a noticeably stronger increase than any analytical theory
proposed so far. As explained in \cite{Bychkov-08}, the stability
problem of a discontinuous deflagration front encounters a deficit
of matching conditions at the front: the number of unknown values
exceeds the number of matching equations by one. In order to
overcome the obstacle, different additional matching conditions were
suggested in different papers \cite{Kadowaki, He, Piriz-01,
Piriz-03, Bychkov-08}. Solution to the problem turned to be quite
sensitive to the choice of the extra condition; but the extra
conditions suggested were merely assumptions. This trouble does not
happen in the numerical solution, since the numerical solution
considers continuous transition from heavy cold plasma to light hot
one. For this reason, the numerical solution provides also a test
for the suggested analytical solutions. In this work we are not
going to criticize the previous analytical theories. Instead, we
will try to extract the best ideas suggested so far in the
theoretical papers \cite{Kadowaki, He, Piriz-01, Piriz-03,
Bychkov-08} to obtain the analytical solution reasonably close to
the numerical one. For that purpose we take the basic elements in
the analysis \cite{Bychkov-08} and complement them by an extra
matching condition identical to that of the DL theory of the
incompressible flow \cite{LandauL}. In the dimensional variables the
DL matching condition is

\begin{equation}
\label{eq45}
\tilde {u}_{za} - {\frac{{\partial f}}{{\partial \tau}} } = 0,
\end{equation}

\noindent where $f$ is perturbation of the discontinuous front
position. The same type of matching condition was assumed for the DL
instability in a compressible deflagration/ablation flow in
\cite{Kadowaki, Piriz-01, Piriz-03}. Reproducing calculations of
\cite{Bychkov-08} with the extra condition (\ref{eq45}) we
obtain the following equation for $\Gamma $

\begin{equation}
\label{eq46} {\frac{{1 - \left( {2\Theta - 1} \right) M\!a_{a}^{2}}}
{{1 - \Theta M\!a_{a}^{2} }}}{\frac{{\Gamma \eta _{c} - 1}}{{\Gamma
- \eta _{c}}} }\left( {\Theta \Gamma + \eta _{a}}  \right) + \Gamma
\eta _{a} + 1 - {\frac{{\Theta - 1}}{{\Theta \Gamma}} }{\frac{{\eta
_{a} + \Theta ^{2}M\!a_{a}^{2} \Gamma ^{3}}}{{1 - \Theta
M\!a_{a}^{2}}} } = 0
\end{equation}

\noindent where

\begin{equation}
\label{eq47} \eta _{a} = \sqrt {1 + M\!a_{a}^{2} \left( {\Theta
^{2}\Gamma ^{2} - 1} \right)} , \quad \eta _{c} = \sqrt {1 +
M\!a_{c}^{2} \left( {\Gamma ^{2} - 1} \right)}
\end{equation}

\noindent and the Mach number in the cold plasma

\begin{equation}
\label{eq48} M\!a_{a}^{2} = {\frac{{M\!a_{c}^{2}}} {{\Theta + \gamma
\left( {\Theta - 1} \right)M\!a_{c}^{2}}} }.
\end{equation}

\noindent Influence of the  Mach number may be illustrated in the limit of small plasma compression, $Ma << 1$, using Taylor  expansion
of Eq. (\ref{eq46}). In that case

\begin{equation}
\label{eq49} \Gamma = \Gamma_0 \left( 1+\beta M\!a_c^2 \right),
\end{equation}

\noindent where $\Gamma_0$ corresponds to DL solution for
incompressible case and $\beta$ is determined as

\begin{equation}
\label{eq50} \beta = 1 - {\frac{{2\Theta \left( {\Gamma _{0} \left(
{\Theta + 2} \right) + 1} \right)}}{{\left( {\Theta + 1}
\right)^{2}\left( {\Gamma _{0} \left( {\Theta + 1} \right) + \Theta}
\right)}}}>0.
\end{equation}

\noindent Positive factor $\beta$ indicates increase of the instability growth rate with plasma compression. The numerical solution to Eq. (\ref{eq46}) together with
Eq. (\ref{eq49}) are presented in Fig. 6 by the solid and dashed
lines respectively. These lines do not provide a perfect agreement
with the numerical results; still the difference between the theory
and the numerical solution is acceptable, about 15\% . For
comparison, Fig. 6 shows also the instability growth rate predicted
in \cite{Kadowaki}, by dash-dotted lines. The theory in
\cite{Kadowaki} differs much stronger from the numerical solution,
approximately by 35\% .

Figure 7 shows the maximal instability growth rate versus the Mach
number for the expansion factors $\Theta = 6,\;10$. As we can see,
in the case of laser ablation with $M\!a_{c} = 1/\sqrt {\gamma}  $
the maximal instability growth rate is about three times larger than
in the incompressible case. These results agree well with the
previous numerical calculations of Ref. \cite{Travnikov-97} for
flames in a compressible flow. It is interesting that the maximal
growth rate shows only minor dependence on the Mach number for a
rather wide range of this parameter, $M\!a_{c} < 0.5$. The strong
dependence of $\sigma _{\max}$ on $M\!a_{c}$ takes place only when
the Mach number starts approaching the limiting value $M\!a_{c} =
1/\sqrt {\gamma}  $ inherent to laser ablation. We observe a similar
tendency in Fig. 8, which presents the cut-off wave number versus
the Mach number. Again, the cut-off wave number is about twice
larger for laser ablation in comparison with the incompressible case
of zero Mach number. Figure 9 compares the cut-off wavelength found
numerically for different values of the Mach number to the
theoretical prediction Eq. (\ref{eq44}). We remind that Eq.
(\ref{eq44}) follows from the theory \cite{Searby and Clavin} in the
limit of an incompressible flow and thermal conduction depending on
temperature as $\kappa \propto T^{5 / 2}$. As we can see, the theory
(\ref{eq44}) provides a reasonable prediction for the cut-off
wavelength in the case of zero Mach number; the difference between
the theory and the numerical solution is about (15-25)\% . As we
increase the Mach number, the DL instability becomes stronger and
the cut-off wavelength decreases considerably. For example, taking
the expansion ratio $\Theta = 7$ we find the cut-off wavelength
$\lambda _{cut} \approx 2.4L_{c} $ for laser ablation, $\lambda
_{cut} \approx 4.8L_{c} $ for the incompressible case of $M\!a_{c} =
0$ and $\lambda _{cut} \approx 6L_{c} $ predicted by the analytical
formula (\ref{eq44}). Thus, the numerical solution predicts the DL
instability in laser ablation on the length scales larger by the
factor of 2.4 than the distance from the ablation zone to the
critical zone. These length scales are very small when compared to
the respective ratio $\lambda _{cut} / L_{c} \approx 20 $ for usual
flames. However, these length scales are extremely large in
comparison with the length scales typical for the RT
instability in inertial confined fusion \cite{Bychkov-94, Betti-95,
Sanz et al.}. For this reason, in order to observe the DL
instability experimentally one has to take special precautions
eliminating the RT instability in the flow. One of the possible
options is to use a sufficiently large target of the thickness
exceeding the distance $L_{c} $ at least by an order of magnitude.
In that case target acceleration is minor, which leads to the
negligible RT instability, while the DL instability has sufficient
space to develop. In addition, one has to remember that experimental
observations typically concern the fastest growing perturbations of
the wavelength $\lambda _{\max}  $ and larger, not the cut-off
wavelength $\lambda _{cut} $. In the case of usual flames these two
length scales are related as $\lambda _{\max}  \approx 2\lambda
_{cut} $, since the whole dispersion relation may be described with
a good accuracy by two first terms in Taylor expansion in
$kL_{c}<<1$, see Eq. (\ref{eq43}). The ratio $\lambda _{\max}  /
\lambda _{cut} $ becomes somewhat different for the DL instability
in laser ablation. Figure 10 shows the ratio $\lambda _{\max}  /
\lambda _{cut} $ obtained numerically for different values of the
Mach number, and compares it to the classical case of $\lambda
_{\max}  / \lambda _{cut} = 2$. We can see that the wavelength
corresponding to the maximal instability growth rate is about
$\lambda _{\max}  \approx 1.8 \lambda _{cut} $ in laser ablation.
The deviation indicates that next order terms become important in
the Taylor expansion of the instability growth rate in  $kL_{c}<<1$.
Still, the deviation is not too large.

Finally, we have to check that our model for the energy gain in the
deflagration/laser ablation flow does not influence the physical
results obtained. In order to validate the model we investigated
influence of the parameters $\beta $ and $n$ of the energy gain on
the DL dispersion relation. Numerical calculations for $\beta = 90$
and different values of $n$ show negligible variations of all
parameters of the instability: the $\Gamma $-factor, the maximal
instability growth rate and the cut-off wavelength. We also took $n
= 2$ and varied $\beta $ within the limits between 20 and 140. For
example, taking $M\!a_{c} = 0.65$ we find the maximal growth rate
$\sigma _{\max}  L_{c} / U_{a} = 0.74$ for $\beta = 20$, $\sigma
_{\max}  L_{c} / U_{a} = 0.84$ for $\beta = 90$ and $\sigma _{\max}
L_{c} / U_{a} = 0.86$ for $\beta = 140$. These calculations indicate
that the continuous numerical model for the energy gain function
$\Omega _{R} $ in Eq. (\ref{eq8}) brings inaccuracy of only few per
cent, about 3\% , into the numerical solution for $\beta = 90$ used
in the present paper. Investigation of the cut-off wavelength for
different $\beta $ and $n$ leads to similar conclusions.

\begin{figure}
\includegraphics[width=.9\columnwidth]{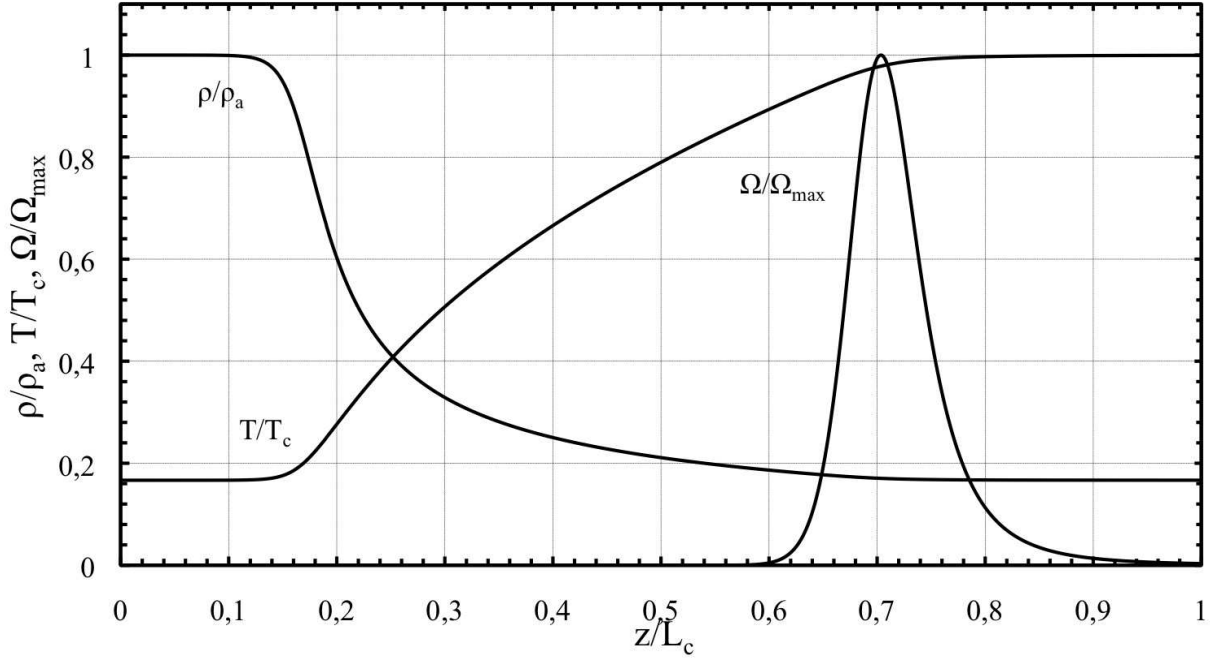}
\caption{Profiles of density, temperature and energy release for
$\Theta= 6$, $M\!a_{c} = 0$, $\beta = 90$.}
\end{figure}

\begin{figure}
\includegraphics[width=.9\columnwidth]{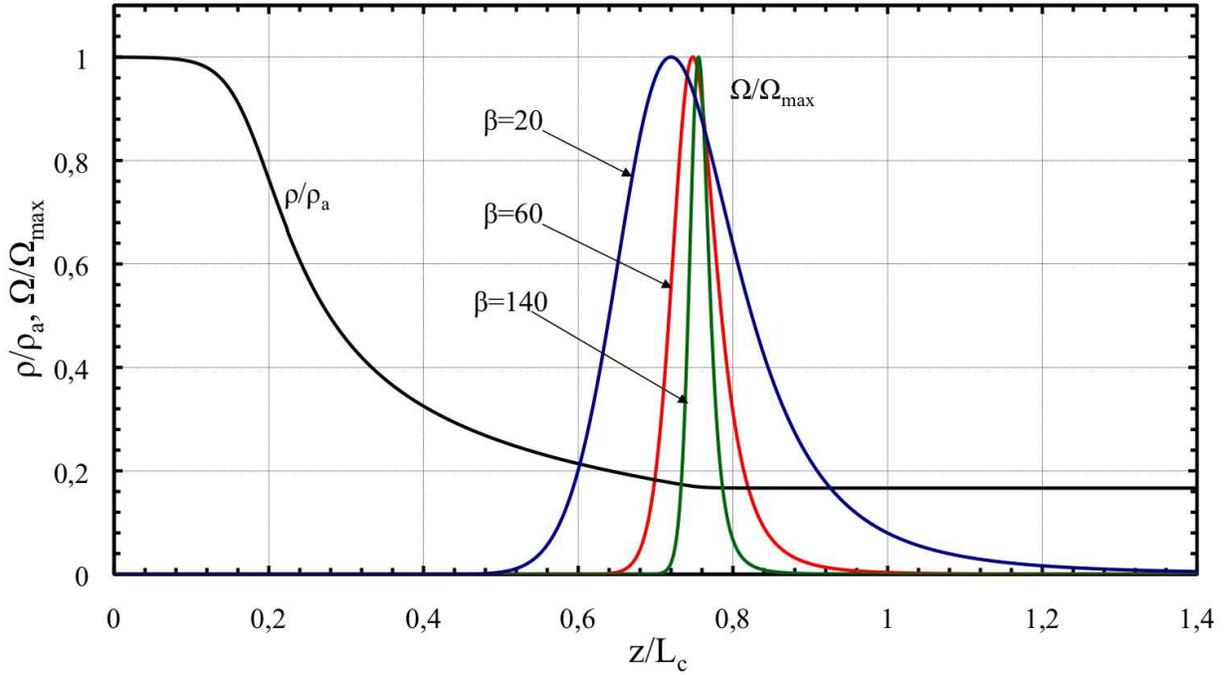}
\caption{Profiles of density and energy release for $\Theta = 6$,
$M\!a_{c} = 0.5$, $\beta = 20;\;60;\;140$.}
\end{figure}

\begin{figure}
\includegraphics[width=.9\columnwidth]{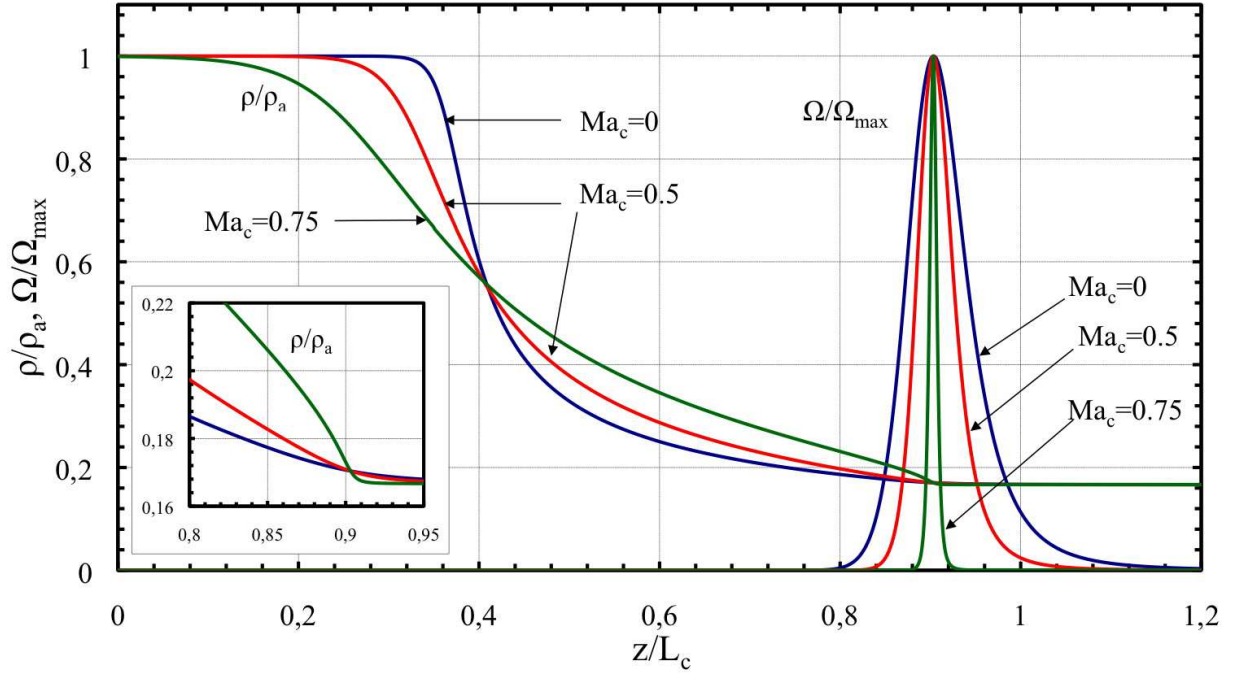}
\caption{Profiles of density and energy release for different Mach
numbers $\Theta = 6$, $M\!a_{c} = 0;\;0.5;\;0.75$.}
\end{figure}

\begin{figure}
\includegraphics[width=.9\columnwidth]{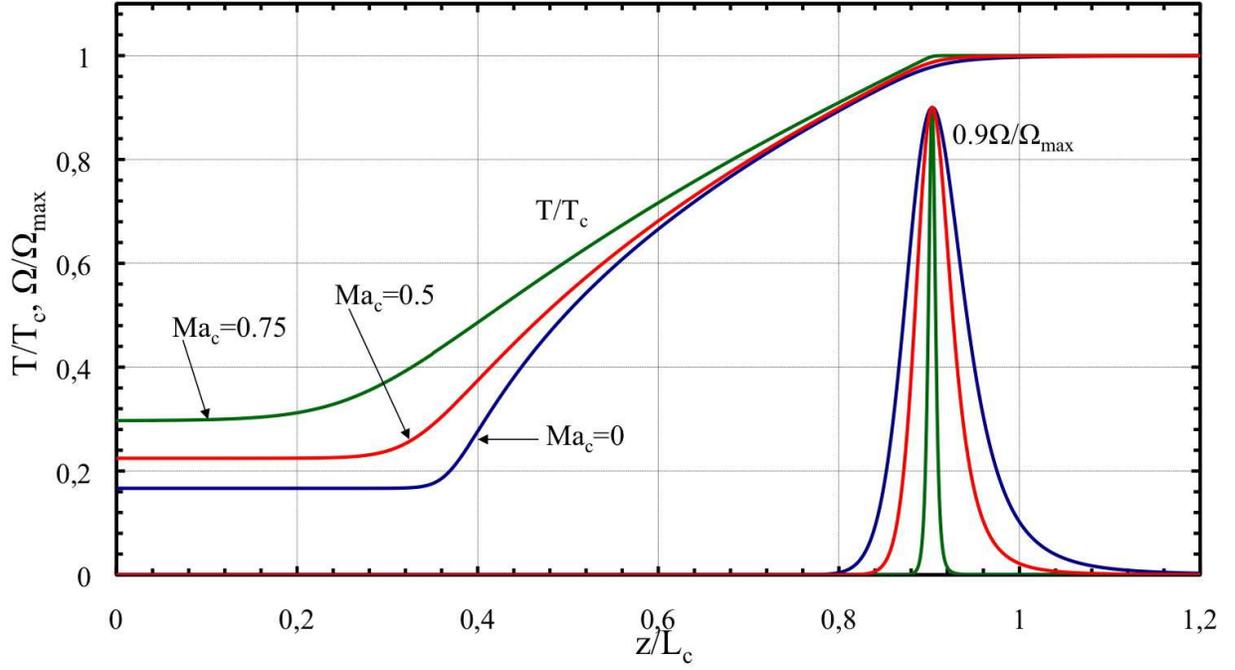}
\caption{Profiles of temperature and energy release for different
Mach numbers for $\Theta = 6$, $M\!a_{c} = 0;\;0.5;\;0.75$.}
\end{figure}

\begin{figure}
\includegraphics[width=.9\columnwidth]{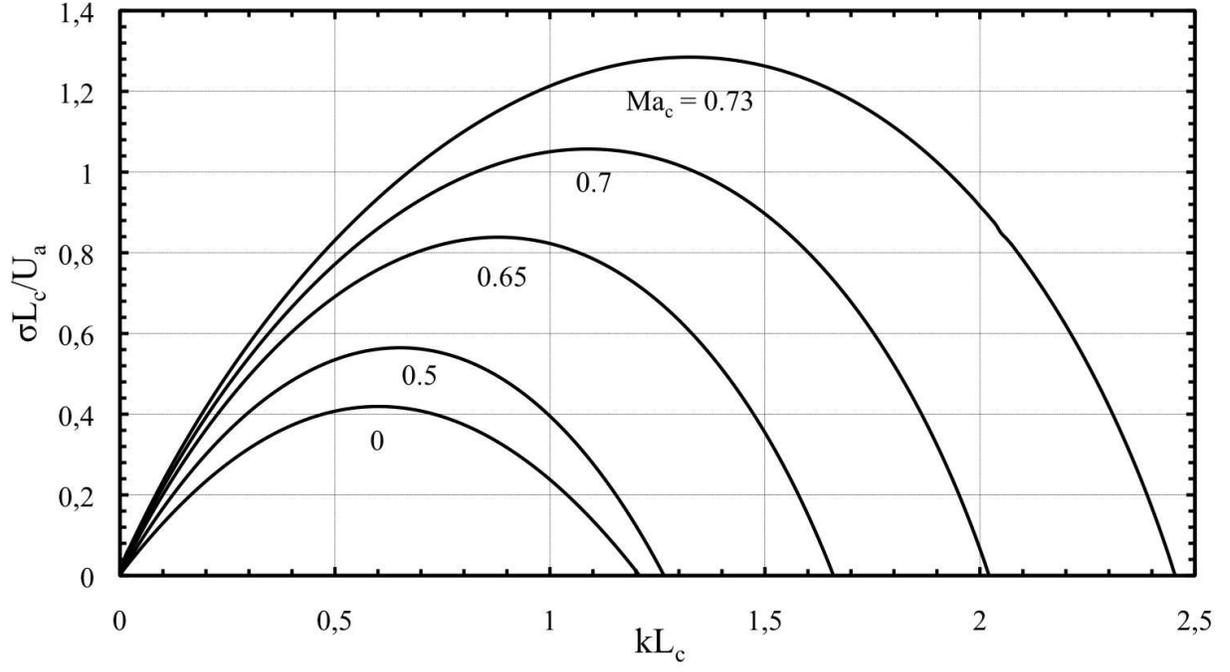}
\caption{Scaled instability growth rate versus the scaled wave number
for different Mach numbers, $\Theta = 6$.}
\end{figure}

\begin{figure}
\includegraphics[width=.9\columnwidth]{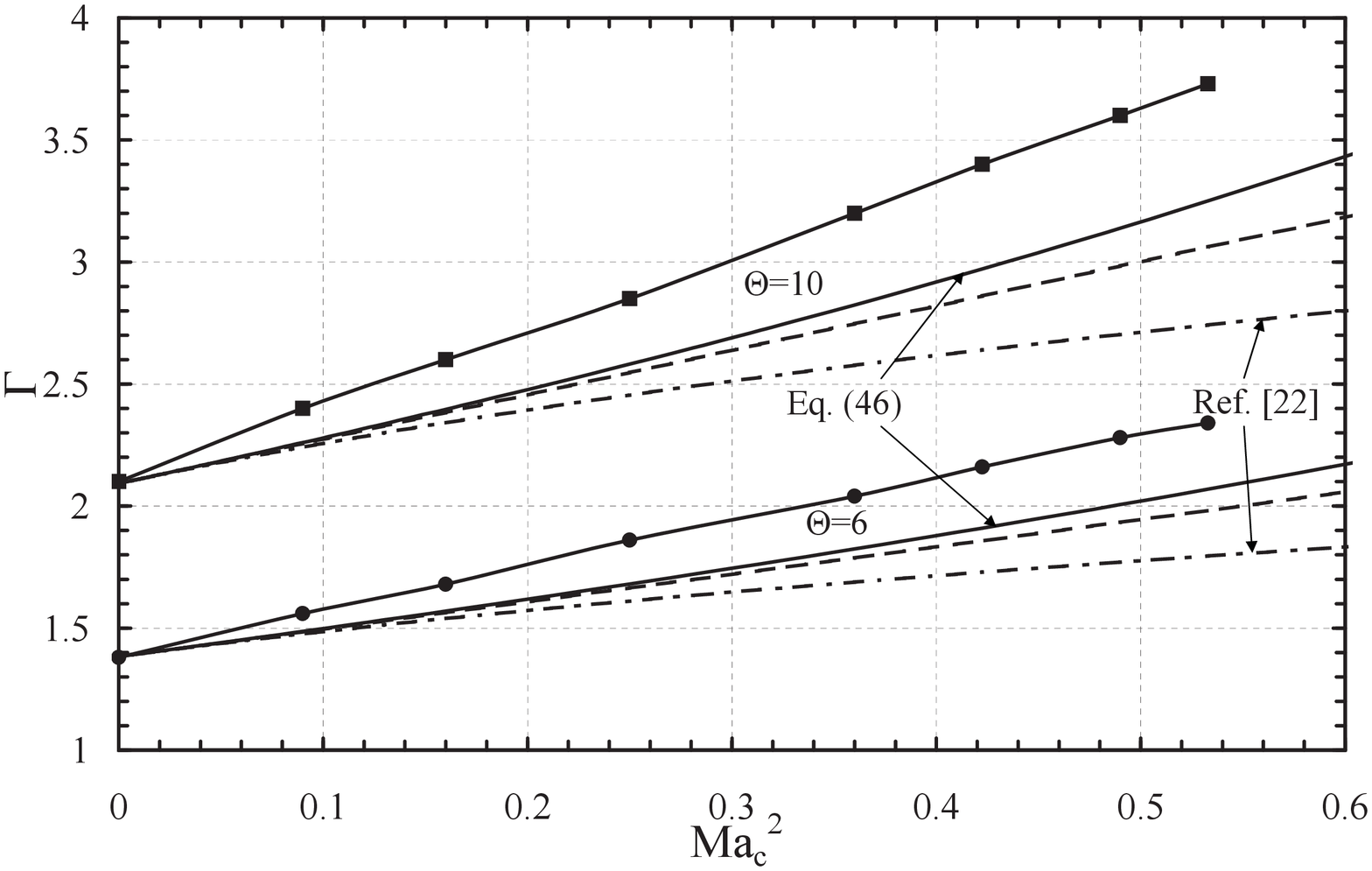}
\caption{The $\Gamma $ factor versus the Mach number squared for
$\Theta = 6$ (circles) $\Theta = 10$ (squares). The solid lines
depict Eq. (\ref{eq46}), the dashed lines corresponds to Eq.
(\ref{eq49}), the dash-dotted lines present the result of Ref.
\cite{Kadowaki}.}
\end{figure}

\begin{figure}
\includegraphics[width=.9\columnwidth]{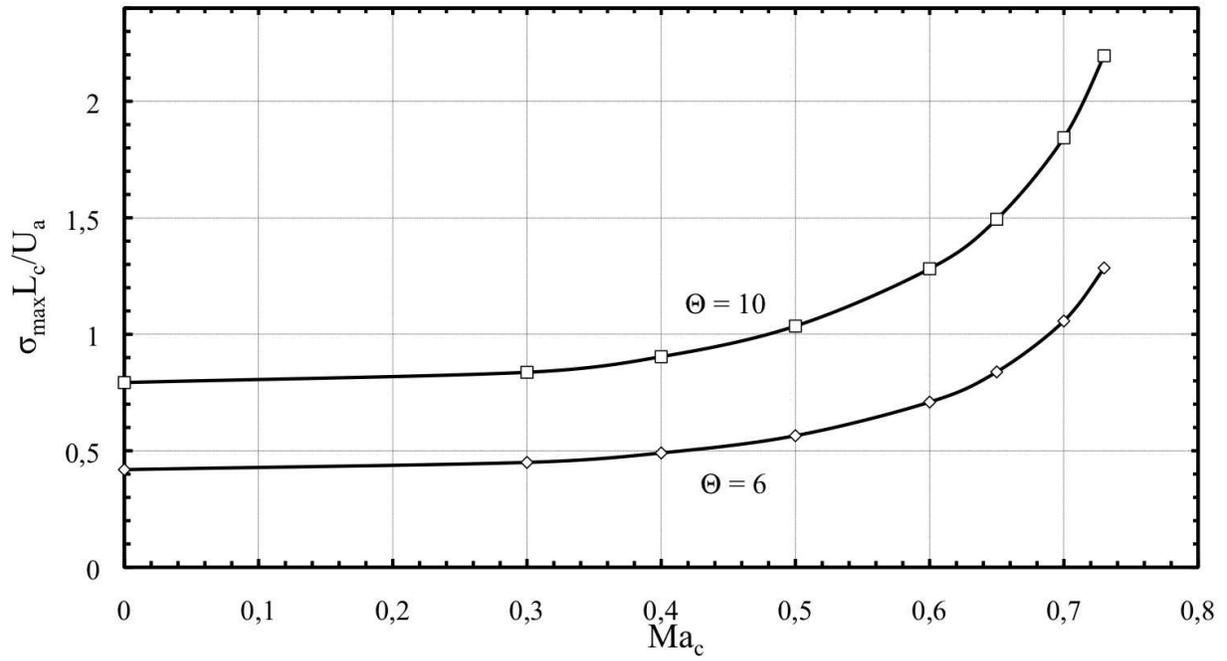}
\caption{Maximum of scaled instability growth rate versus the Mach
number for $\Theta = 6;\;10$.}
\end{figure}

\begin{figure}
\includegraphics[width=.9\columnwidth]{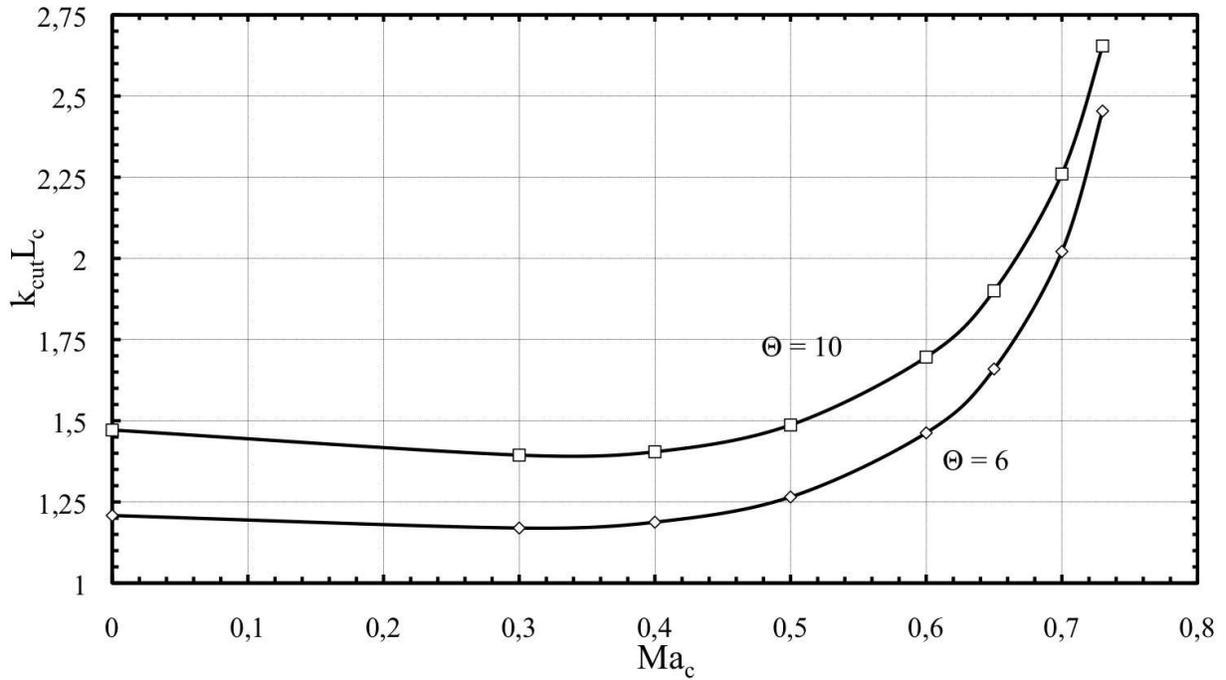}
\caption{The cut off wave number versus the Mach number, $\Theta =
6;\;10$.}
\end{figure}

\begin{figure}
\includegraphics[width=.9\columnwidth]{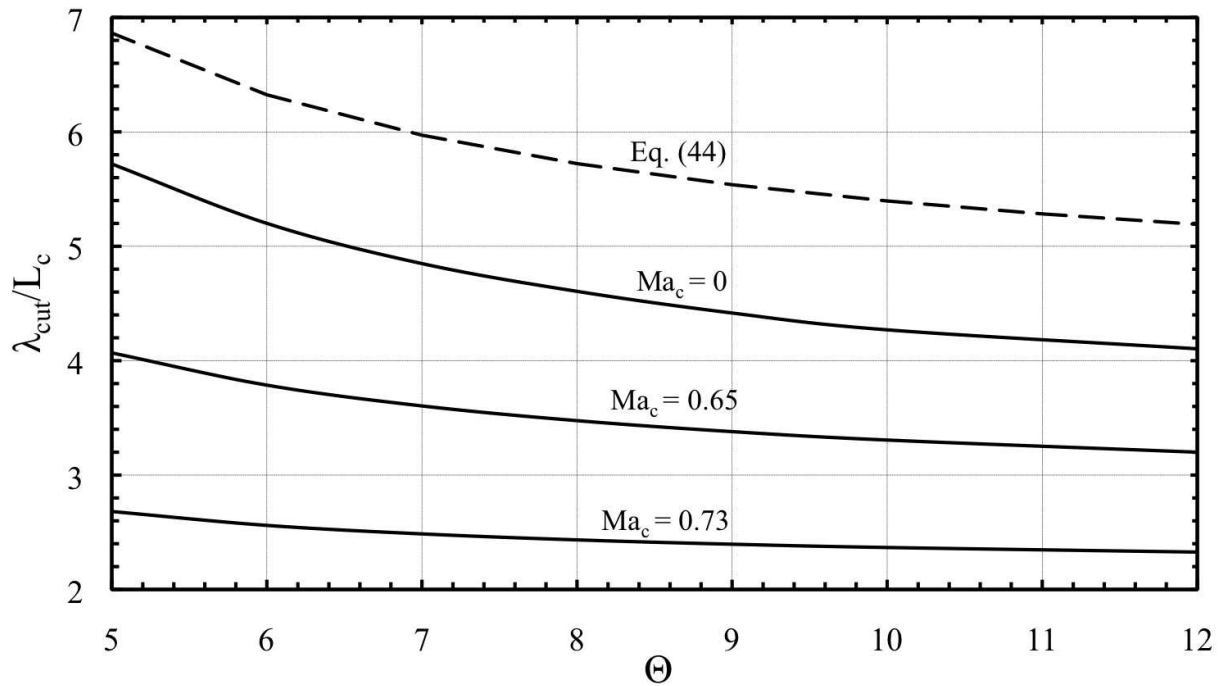}
\caption{The cut off wavelength versus the expansion factor for
$Ma_{c} = 0;\;0.65;\;0.73$. The dashed line shows the analytical
formula Eq. (\ref{eq44}).}
\end{figure}

\begin{figure}
\includegraphics[width=.9\columnwidth]{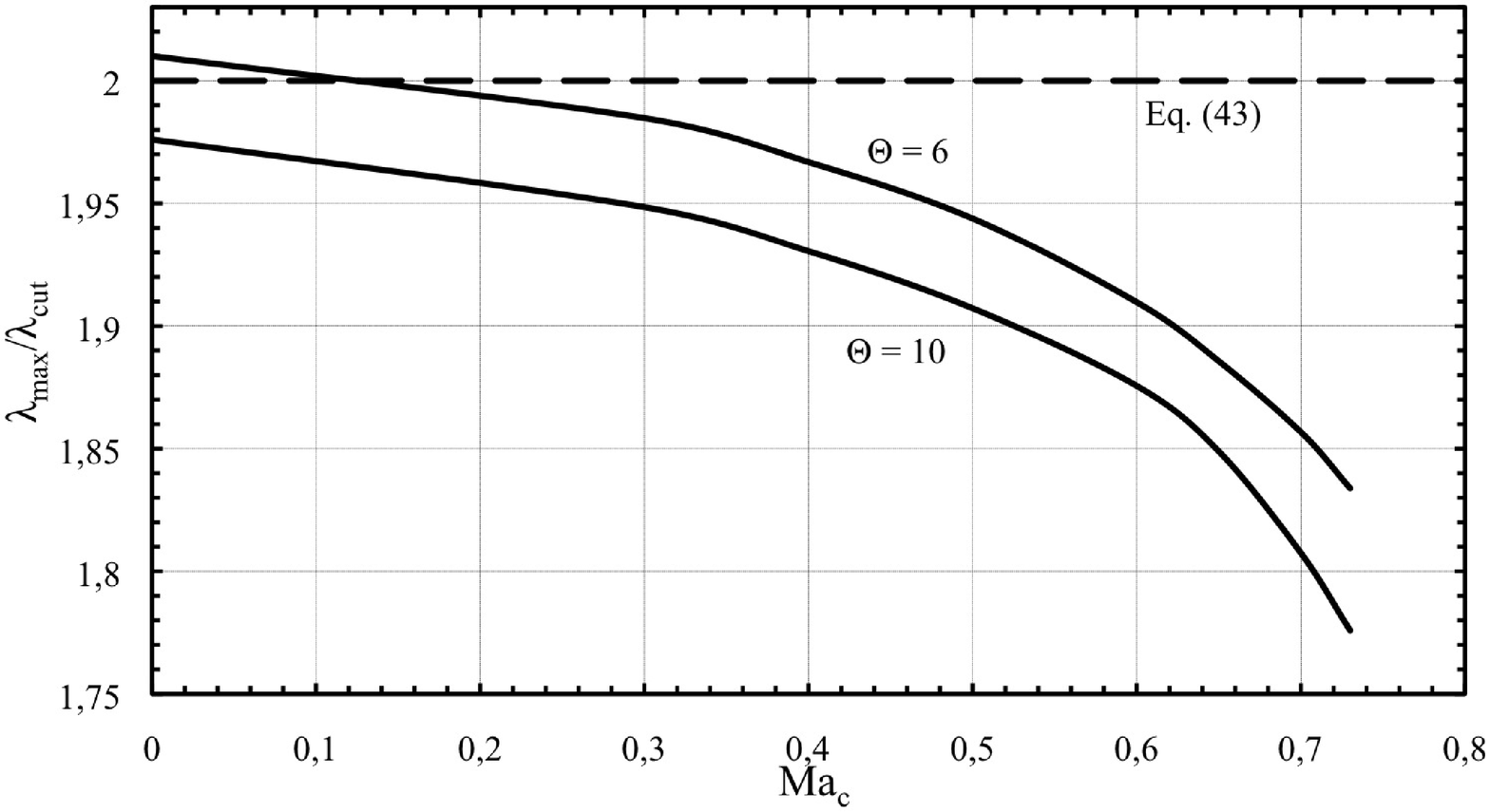}
\caption{Ratio of the wavelength of maximal growth rate and the
cut-off wavelength $\lambda _{\max}  / \lambda _{cut} $ versus the
Mach number for $\Theta = 6;\;10$. The dashed line shows the
theoretical prediction $\lambda_{max} / \lambda_{cut} = 2$, Eq.
(\ref{eq43}).}
\end{figure}

\section{Conclusions}

In the present paper we investigated the DL instability in an
ablation flow and compared the results to the classical case of a
slow flame. Unlike the normal flame, laser ablation is
characterized by the strongest plasma compression possible
for a deflagration wave. Another specific feature of laser ablation
is the strong dependence of electron thermal conduction on temperature.
We demonstrate that the DL instability in laser ablation is much
stronger than in the classical case. In particular, the maximal
growth rate in the ablation flow is about three times larger than in
the incompressible case. Moreover, the cut-off wavelength changes drastically
as we go from the classical case of an incompressible flow
to the ablation flow. The cut-off wavelength is also strongly
influenced by the temperature dependence of thermal conduction. It
is known that the DL instability for usual flames develops on quite
large length scales exceeding the flame thickness by almost two
orders of magnitude \cite{review}. In contrast to this, the
characteristic length scale of the DL instability in the ablation
flow (e.g., the cut-off wavelength) is comparable to the total
distance from the ablation zone to the critical zone of laser light
absorption, $L_{c} $. Still, even these values are  large from
the point of view of possible experimental observations of the DL
instability in laser ablation. We note that the RT instability in
inertial confined fusion develops on length scales much smaller
than $L_{c} $. For this reason, the DL instability may be observed
only if the accompanied RT instability is suppressed. This may be
achieved, for example, for sufficiently large targets of thickness
much larger than the distance $L_{c} $ from the critical to the
ablation zone.

\section*{Acknowledgements}

This work was supported by the Swedish Research council and by the
Kempe foundation.




\end{document}